\documentstyle[12pt]{article}
\def\be{\begin{equation}}
\def\ee{\end{equation}}
\def\bea{\begin{eqnarray}}
\def\eea{\end{eqnarray}}
\def\d{\partial}

\def\ncs{N_{\mbox{\scriptsize\rm CS}}}

\def\bm#1{\mbox{\boldmath $#1$}}
\def\esp{E_{{\rm sph}}}
\def\gsim{{\,}^{\raisebox{-.4ex}{${\scriptstyle>}$}}_
{\raisebox{.4ex}{${\scriptstyle\sim}$}}\,}
\def\lsim{{\,}^{\raisebox{-.4ex}{${\scriptstyle<}$}}_
{\raisebox{.4ex}{${\scriptstyle\sim}$}}\,}

\def\aw{\alpha_{\rm w}}
\begin{document}
\baselineskip 22pt
\title{\vskip-1.5truecm{\hfill \baselineskip 14pt {
\small  FERMILAB-Conf-96/266-A}\vskip .1truecm}\vskip.1truecm
{\hfill \baselineskip 14pt {
\small  hep-ph/9608456}\vskip .1truecm}
\vskip 0.3truecm {Baryon Number Non-Conservation and the Topology of Gauge
Fields}}
\author{
Minos Axenides$^{1,}$\thanks{email:
{\tt axenides@talos.cc.uch.gr}}\ ,
Andrei Johansen$^{2,}$\thanks{email:
{\tt johansen@string.harvard.edu}}\ ,
Holger B. Nielsen$^{3,}$\thanks{email:
{\tt hbech@nbivms.nbi.dk}}\\[.4cm] and
Ola T\"{o}rnkvist$^{4,}$\thanks{email:
{\tt olat@fnal.gov}}\\*[1.5cm]
{\normalsize\it $^1$University of Crete, GR-71409 Iraklion, Greece}\\[1ex]
{\normalsize\it $^2$Lyman Laboratory of Physics, Harvard University,
Cambridge, MA 02138, USA }\\[1ex]
{\normalsize\it $^3$The Niels Bohr Institute, Blegdamsvej 17, DK-2100
Copenhagen,
Denmark}\\[1ex]
{\normalsize\it $^4$NASA/Fermilab Astrophysics Center,
Fermi National Accelerator Laboratory,}\\ {\normalsize\it Box 500,
Batavia, IL 60510-0500, USA}}
\def\today{}
\maketitle
\begin{abstract}
\baselineskip 18pt
An introduction to the subject of baryon number
non-conservation in the electroweak theory at high temperatures
or energies is followed by a summary of our discovery
of an infinite surface of sphaleron-like configurations which play
a key role in baryon-number non-conserving transitions in a hot
electroweak plasma.
\end{abstract}
\vspace*{1cm}

\noindent
Talk given by O.T.\ at the meeting of the American Physical Society, Division
of
Particles and Fields (DPF96) in Minneapolis, Minnesota, August 10-15, 1996, to
appear
in the proceedings.

\thispagestyle{empty}

\newpage
\pagestyle{plain}
\setcounter{page}{1}

\section{Introduction}

Any electroweak gauge theory built on
SU(2)$\times$U(1) with chiral fermions,
such as the Standard Model,
permits non-perturbative processes that do not conserve
baryon number ($B$). The processes are associated with an
energy barrier of height $\esp\sim 4 M_W/\aw\approx 10$ TeV
that suppresses the tunneling rate at zero energy and temperature
with a factor $\exp(-4\pi/\aw)\approx e^{-170}$.
The $B$ non-conserving transition rates become unsuppressed in a
high-temperature
electroweak plasma and possibly also in
high-energy particle collisions\cite{ShapRub}.

Within an extended Standard Model the
observed excess of matter versus antimatter in the universe may
be explained as originating from $B$ non-conserving processes
in the hot electroweak phase transition ($T\gsim 100-300$ GeV).
In multiparticle high-energy collisions with a center-of-mass energy
$E$ it has been shown perturbatively that the
tunneling cross-section increases
exponentially in $E^{4/3}$ for $E\ll\esp$, but the perturbative analysis
becomes unreliable for $E\lsim\esp$. Non-perturbative approaches to
barrier tunneling at high-$E$ collisions attract
much interest and require further study.

Because of the chiral anomaly
the baryon number is related to the Chern-Simons number $\ncs$,
which characterizes the winding of the gauge fields, by the conservation
law $\Delta (B - n_{\rm F} \ncs)=0$ where $n_{\rm F}$ is the number
of families. $B$ non-conservation is therefore accompanied by
large, non-perturbative
changes in the gauge fields corresponding to transitions
between degenerate minima of the  SU(2)$\times$U(1) vacuum.
The existence of an infinite number of such minima, labeled by
an integer $\ncs$, is a consequence of the fact that
the SU(2) field tensor
$W^A_{ij}$ (the square of which appears
in the energy) becomes zero whenever $W_i\equiv W^A_i\sigma_A$ is
of the pure-gauge form
\be
\label{pure}
W_i=-i(\d_iU(\bm{x})) U^{-1}(\bm{x}).
\ee
Two adjacent minima $\ncs=n$ and
$\ncs=n+1$, $n\in{\bf Z}$, are separated by a
high energy barrier.
The highest point on the lowest possible path across the barrier
corresponds to the energy $\esp$
and has
been identified as the {\em sphaleron\/} solution\,\cite{KM}.
It is a saddle point in configuration space with only two descending
directions and Chern-Simons number $\ncs = n+ 1/2$.

\section{Baryon number non-conservation in a hot electroweak plasma}
In a hot  electroweak
plasma $\Delta B\neq 0$ processes have been thought to
occur via classical thermal fluctuations across the barrier
through the sphaleron saddle point. For temperatures $T\ll\esp$,
the transition rate $\Gamma$ is proportional to the Boltzmann factor
$\exp(-\esp/T)$. For temperatures in the range $M_W\lsim T
\lsim\esp$,  Gaussian fluctuations about the sphaleron produce
a prefactor\cite{AM} $M_W^4(\esp/ T)^3$.
Because of the prefactor, the transition rate would
decrease steeply at high temperatures\footnote{Recent
calculations of bosonic and fermionic 1-loop corrections
to the transition rate do not affect the prefactor; see Ref.\ 4 and references
therein.}.
This result is in contradiction with a scaling argument suggesting
that $\Gamma\sim T^4$ in the symmetric phase ($T$ being the only
dimensionful parameter) as well as with lattice real-time
simulations\,\cite{AmbKras} that find
$\Gamma=\kappa (\alpha_w T)^4$, $\kappa\approx 1$.

In a recent publication\,\cite{paper}
we have shown the existence of an infinite
``surface'' of configurations with sphaleron-like properties, on
which the sphaleron configuration is the point of lowest
energy. These configurations mediate $B$ non-conserving
processes in the hot broken phase as well as in the symmetric phase.
For temperatures $T\lsim\esp$, the probability of thermal
fluctuations with energy $E>\esp$ is considerable, and
many classical paths other than those leading through the
sphaleron are accessible and contribute to the transition
rate. The perturbative expansion around the
sphaleron is unreliable for $E\gsim\esp$ partly because, in our view,
the multitude of sphaleron-like configurations with energy near
$\esp$ is not properly accounted for in that approximation.

\section{Infinite surface of sphaleron-like
configurations}
Sphaleron-like configurations\,\footnote{For
simplicity we present here only the case of
SU(2) and vanishing Yukawa couplings. For more
general cases, see Ref.\ 6
and references therein.} can be identified through
the following
observations about the sphaleron solution\,\cite{KM}:
(i) its gauge field is
odd, $W_i(\bm{x})=- W_i(-\bm{x})$, and approaches the form
(\ref{pure}) at infinity, where $U(\bm{x})$ is likewise odd,
(ii) it has a half-integer Chern-Simons number $\ncs = n+ 1/2$,
$n\in {\bf Z}$, (iii) the Dirac equation in
the sphaleron gauge-field background has exactly one zero-mode.
For this reason, one fermion energy level crosses zero in a
transition through the sphaleron.
In the Dirac sea picture, this is
consistent with the creation or destruction of a baryon (anti-baryon).
In fact, baryon number non-conservation requires, in addition to
a change in Chern-Simons number, the crossing of an odd number of
fermion energy levels.

The property (i) leads us to consider the class of
{\it generalized odd\/} fields,
\be
\label{genodd}
W_i(-\bm{x})=-W^{S}_{i}(\bm{x})\equiv
-[S(\bm{x})W_{i}(\bm{x})S^{-1}(\bm{x}) +
i\partial_{i}S(\bm{x})S^{-1}(\bm{x})],
\ee
for some element $S$ of the gauge group. This class of fields includes
the odd fields but is much larger. The definition (\ref{genodd})
of generalized oddness is gauge invariant. All results presented
below have been obtained for generalized odd fields\,\cite{paper},
but for simplicity we restrict to ordinary odd fields in this
presentation.

{}From property (i) and Eq. (\ref{pure}) one easily derives that
$\d_i[U^{-1}(-\bm{x})U(\bm{x})]=0$, and thus
$U(\bm{x})=\pm U(-\bm{x})$ as $|\bm{x}|\to\infty$. The class of
odd fields therefore naturally splits into two disconnected
classes, for
which one can show the following: Fields with even $U$ at infinity
have integer $\ncs$ and are continuously connected to a minimum of
the vacuum through odd-parity fields. Fields with odd $U$ at
infinity have half-integer $\ncs$ and are continuous
deformations
of the sphaleron within the class of odd-parity fields\,\footnote{These
results have
an independent proof within the theory of homotopy groups of
maps\,\cite{paper}.}.

The  Dirac equation in a background of odd-parity
fields $W_i$ with odd $U$ can be shown to have an odd number
of zero-modes. Therefore,
a transition through any such sphaleron-like configuration
will lead to the
crossing of an odd number of fermion energy levels.
The proof is simple: Consider
an eigenfunction $\psi(\bm{x})$ to the three-dimensional Dirac
equation $\sigma_i(i\d_i - W_i)\psi(\bm{x}) = E \psi(\bm{x})$
with energy $E\neq 0$. Then $\psi(-\bm{x})$ is an eigenfunction
corresponding to the energy eigenvalue $-E$. Thus, non-zero eigenvalues
are paired ($E$,$-E$).
Consider now a continuous deformation of the gauge field away
from the sphaleron, in whose background field we know that the Dirac
equation has one zero-mode. As the field is varied within
the class of odd-parity fields,
positive and negative eigenvalues will appear or disappear
in pairs, and the number of zero-modes will stay odd.

Although there is no reason to believe that all $B$
non-conserving
processes occur in transitions through
odd-parity odd-U fields, there is an argument which suggests that
such configurations are energetically favored in thermal
fluctuations near and above $\esp$. Put in other
words, the energy rises steeply to inaccessible values as one
ascends from the sphaleron
in all directions except along the odd-parity configurations. The
simple reason is that the energy functional is invariant when
$W_i(\bm{x})\to - W_i(-\bm{x})$ for an arbitrary field $W_i$,
and the odd-parity fields constitute a fixpoint under this
transformation.

\section{Conclusions}
We find an infinite set of configurations other than the sphaleron
which are the loci of fermion energy-level crossings
in baryon number non-conserving thermal transitions.
This result is independent of
the Higgs sector and applies equally to the broken and the symmetric
phases. The configurations, odd under a generalized
parity, are easily excited thermally for $T\lsim\esp$ and play
a key role in baryon number non-conserving transitions in a hot
electroweak plasma. Their relevance in high-energy collisions remains to
be investigated.

\section*{Acknowledgments}
Support for M.A. provided by Danmarks Grundforskningsfond and the European
Union contract no.\ ERBCHRXCT $940621$; for A.J.\ by NATO
grant GRG $930395$; for O.T.\ by DOE and NASA under Grant NAG$5$--$2788$
and by the Swedish Natural Science Research Council (NFR).

\end{document}